\documentclass[journal]{IEEEtran}

\usepackage{graphicx}
\usepackage{cite}
\usepackage{multirow}
\usepackage[cmex10]{amsmath}

\usepackage{cite}

\usepackage{multirow}
\usepackage{multicol}
\usepackage{mathtools}

\usepackage{amsmath,epsfig}
\usepackage{amsfonts}
\usepackage{url}
\usepackage{amssymb}
\usepackage{amsthm}

\usepackage{caption}
\usepackage{subcaption}

\begin{document}
%
\title{Semi-supervised learning for joint SAR and multispectral land cover classification}
%
%
%

\author{Antonio~Montanaro, Diego~Valsesia, Giulia~Fracastoro,
        and~Enrico~Magli
\thanks{The authors are with Politecnico di Torino, Torino, Italy. Code is available online: https://github.com/diegovalsesia/sscl}}

\maketitle

\begin{abstract}
Semi-supervised learning techniques are gaining popularity due to their capability of building models that are effective, even when scarce amounts of labeled data are available. In this paper, we present a framework and specific tasks for self-supervised pretraining of \textit{multichannel} models, such as the fusion of multispectral and synthetic aperture radar images. We show that the proposed self-supervised approach is highly effective at learning features that correlate with the labels for land cover classification. This is enabled by an explicit design of pretraining tasks which promotes bridging the gaps between sensing modalities and exploiting the spectral characteristics of the input. In a semi-supervised setting, when limited labels are available, using the proposed self-supervised pretraining, followed by supervised finetuning for land cover classification with SAR and multispectral data, outperforms conventional approaches such as purely supervised learning, initialization from training on ImageNet and other recent self-supervised approaches. 
\end{abstract}


%
\IEEEpeerreviewmaketitle

\section{Introduction}
Deep learning is nowadays an established way of designing powerful models that are able to effectively solve problems in a wide variety of fields, from natural language processing, to computer vision and remote sensing. The most striking successes are obtained by supervised learning, where huge annotated datasets are used to learn end-to-end models addressing a specific task. However, supervised learning has been increasingly under scrutiny due to data requirements, since huge datasets, like ImageNet, are not available in all domains. This is the case of remote sensing imagery, where carefully annotating satellite images requires domain experts, and doing so for large amounts of data can be expensive and error-prone. 

The emerging field of self-supervised learning (SSL) addresses this data bottleneck, studying techniques that can be used to train deep models to extract features that are relevant to the problem of interest, without requiring labeled data.

This paper addresses the problem of developing SSL techniques that are effective for the land cover classification problem in remote sensing. This is not a trivial objective since there are several challenges that are unique to this problem and find no correspondence in other fields such as the computer vision field. In particular, in Earth observation, several imaging modalities (e.g., optical and radar) can be used to acquire a scene of interest, and it is not obvious how to train a model that is capable of exploiting both. In this paper we address the problem of using multiple imaging modalities, namely multispectral and synthetic aperture radar (SAR) images, to infer the land cover classes, proposing a general and modular framework that does not pose specific requirements on the employed neural network architecture.

Recent works in the context of the 2020 IEEE GRSS Data Fusion Contest \cite{9369830} have shown difficulties in building competitive end-to-end models based on deep learning for land cover classification with both SAR and multispectral data. 
This is a symptom of deep models being unable to extract high-quality features due to a variety of reasons such as difficulties in integrating two widely different imaging modalities, lack of large labeled datasets, pretraining techniques suffering from large domain gaps with respect to remote sensing data, and more. 

For this reason, we propose a method, named Spatial-Spectral Context Learning (SSCL), which is composed of a generic modular architecture for neural networks and two self-supervised pretraining approaches, allowing to effectively train models for multichannel data having an arbitrary number of channels representing imaging modalities (multispectral bands, SAR polarizations, etc.). SSCL is a universal framework that can be used whenever the available input data have many channels and it is more effective than transfering models from computer vision datasets due to the large existing domain gaps. For example, image classification on ImageNet deals with RGB instead of multichannel images, its classes are mostly object-centric and require reasoning about spatial geometry rather than spectral characteristic of materials. Instead, the self-supervised tasks in SSCL are explicitly designed to account for the existence of multiple channels with possibly very different representations, and promote learning a model of the correlations across channels, as the spectral properties of materials can be jointly inferred from the visible and infrared spectral bands in multispectral images, and from the microwave wavelengths captured by SAR. Since the classes of interest in problems such as land cover classification involve discriminating materials, this multichannel approach is more effective at extracting features for remote sensing problems.

Extensive experiments show how the proposed method is effective in the semi-supervised setting, where the model pretrained with self-supervision is finetuned with a few labels. In particular, SSCL is superior to purely supervised learning, pretraining from ImageNet and recent self-supervised pretraining paradigms from computer vision \cite{chen2021exploring} and remote sensing \cite{chen2021self}, when labels are scarce.

\section{Related work}
\label{sec:background}
Recently, many researchers have started investigating SSL approaches since they do not require external labelled data.
The most popular approach consists in learning to capture relevant image features by solving a pretext task. 
A wide variety of pretext tasks have been proposed \cite{jing2020self}.
Some of them involve geometric transformations such as guessing the rotational angle of an image, others consider generation-based tasks such as image inpainting. 
More recently, contrastive learning is emerging as a new appealing paradigm for SSL.
This approach aims at embedding augmented views of the same input close to each other, while trying to separate embeddings from different inputs.
All the methods following this approach employ a siamese network and a contrastive loss \cite{jaiswal2021survey}, but they differ from each other mainly in the way they collect negative samples.

Remote sensing is strongly affected by limited data availability, where datasets are several but sparsely annotated.
In order to overcome these issues, a limited number of works have started to explore using SSL approaches in remote sensing applications, in particular for land scene classification.
In \cite{9413112}, the authors propose to use colorization as pretext task for remote sensing imagery, leveraging the spectral bands to recover the visible colors. 
Instead, in \cite{tao2020remote} the authors compare three different SSL techniques, namely image inpainting, relative position prediction and instance discrimination, showing that the latter provides better performance for scene classification.
Another work \cite{saumoco} extends the constrastive approach proposed by MoCo to remote sensing imagery, defining the augmented views as randomly shifted patches of the same image. 

However, little attention has been paid to develop self-supervised deep learning models that can effectively combine information from different spectral channels or sensing modalities, such as multispectral and SAR. In this field, the most common techniques are still based on standard machine learning methods. Most of them are supervised methods \cite{sathya2017analysis}, \cite{lo2004hybrid}, and few are unsupervised \cite{abbas2016k}. Contrastive Multiview Coding (CMC) \cite{stojnic2021self} tries to combine information from different channel subsets of a multispectral image, using a contrastive approach. Although this method seems to be effective when evaluated using a linear classification protocol after SSL only, it is not able to improve over the classic ImageNet pretraining in the semi-supervised setting, when supervised finetuning is performed. This might be a symptom that the features learned via SSL do not generalize well and supervision has to undo part of the learning process. In addition, it does not consider land cover mapping as downstream task. Recently, Chen et al. \cite{chen2021self} proposed SSL for joint land cover classification with SAR and multispectral images adopting a contrastive approach at image level and super-pixel level. As shown in Sec. \ref{sec:main_results} can be considered complementary to our work, as it is superior in the self-supervised regime, while SSCL outperforms it in the semi-supervised finetuning regime.

\vspace{-4pt}
\section{Proposed method}

\begin{figure*}
    \vspace*{-4pt}
    \begin{subfigure}[b]{0.22\textwidth}
    \centering
    \includegraphics[width=0.99\textwidth]{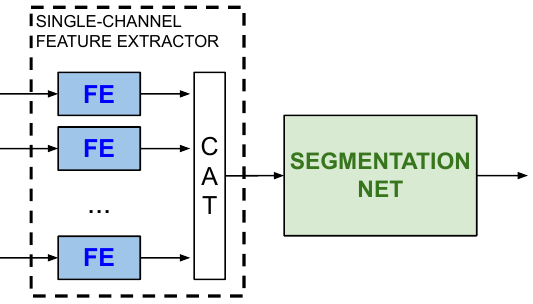}
    \caption{Architecture}
    \label{fig:architecture}
    \end{subfigure}
    \vspace*{-4pt}
    \begin{subfigure}[b]{0.35\textwidth}
    \centering
    \includegraphics[width=0.99\textwidth]{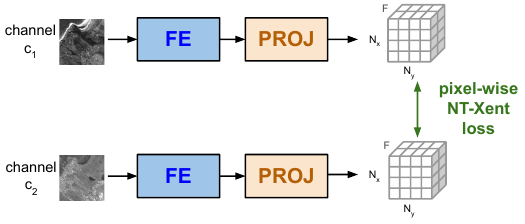}
    \caption{UniFeat}
    \label{fig:unifeat}
    \end{subfigure}
    \begin{subfigure}[b]{0.43\textwidth}
    \centering
    \includegraphics[width=0.99\textwidth]{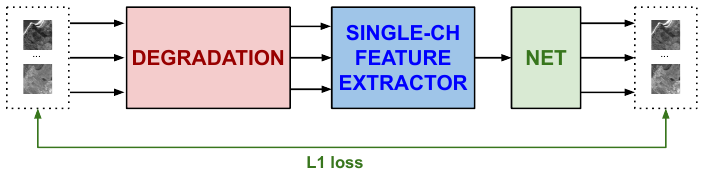}
    \caption{CoRe}
    \label{fig:core}
    \end{subfigure}
    \vspace*{-10pt}
    \caption{General architecture and self-supervised pretraining stages. a) Overall architecture: each channel of the input is processed independently by the same feature extractor (FE) via weight sharing. Outputs are concatenated along the feature axis and fed to a state-of-the-art network for image segmentation; b) \textit{Unifeat}: contrastive learning pretrains the single-channel FE to bring features of different sensing modalities closer; c) \textit{CoRe}: Context Reconstruction from dropped channels, spatial areas and blur pretrains the entire architecture to promote feature clustering according to spectral material properties.}
\vspace*{-10pt}
\end{figure*}

In this section we present the proposed approach to land cover mapping from joint SAR and multispectral imagery, i.e. Spatial-Spectral Context Learning (SSCL).

The main novelty of the proposed method lies in the development of self-supervised pretraining strategies that are able to train feature extractors for the land cover classification task. If labeled data are available, further supervised finetuning can be performed to achieve improved performance.

The proposed self-supervised approach comprises two stages of pretraining, which we call \textit{Unifeat} and \textit{CoRe}, accomplishing different goals. In addition, an important concept that we introduce regards the overall neural network architecture, which is illustrated in Fig. \ref{fig:architecture}. State-of-the-art semantic segmentation models are often developed for single-band or RGB images. It is important to carefully adapt them to the scenario where multiple channels, possibly from multiple imaging modalities, are available. For this reason, we also present  a preprocessing stage, composed by a few convolutional layers, acting on the individual channels and sharing its model weights across them. The goal is to slowly extract features from the single channels themselves, before merging them. We call this block as single-channel Feature Extractor (single-channel FE).
This, compared to early fusion, allows to build a richer feature space and ties into the working of the first stage of self-supervised pretraining, which promotes a convergence of the statistics of the various channels to reduce their domain gap.  It is also a flexible approach that can be used for any number of spectral bands or sensing modalities.

\label{sec:method_selfsup}
\subsubsection{UniFeat -- contrastive uniforming of sensing modalities}
A first issue lies in the multi-channel nature of the input and the domain gap that exists between the channels, particularly different sensing modalities such as SAR and optical images due to coherent and incoherent imaging.
Since the same scene is being imaged across the modalities, it is desirable for the features that are derived to be robust to low-level variations which do not carry discriminative information to infer the class label. Examples of such low-level nuisances can be the different noise characteristics of each channel, the local patch statistics, and so on. Promoting similarity of low-level features across the input channels can help bridge the domain gaps, and avoid large distances between points in the feature space representing the same class.
This is the goal of the first self-supervised task we propose, namely \textit{UniFeat}, depicted in Fig. \ref{fig:unifeat}. This task addresses the pretraining of the single-channel FE.
We consider the features extracted by the single-channel encoder, consisting in one vector with $F$ features for each spatial location $(i,j)$ and each channel $c$. We use a contrastive learning approach where we promote similarity between the feature vectors of two patches representing the same area from different input channels. Conversely, dissimilarity is promoted if the patches do not represent the same geographical area. Several contrastive losses have been studied for this kind of tasks in computer vision problems \cite{jaiswal2021survey}. We choose to follow the SimCLR approach \cite{chen2020simple}, where we consider the single-channel feature extractor as the base encoder $f(\cdot)$ and we introduce an additional projection head $g(\cdot)$ that maps the output features of the single-channel encoder to the space where the discriminative loss is applied. Notice that, contrary to the base encoder adopted in SimCLR which targets whole-image classification, the proposed single-channel encoder does not pool all the feature vectors of the patch into a single representation to be further projected, but rather produces a pixel-wise mapping of the input. 
This promotes features with higher spatial resolution, as shown in Sec. \ref{sec:results}, which is particularly useful for the land cover classification task. 
The projection head depicted in Fig.\ref{fig:unifeat} is removed after pretraining.

More in detail, given a minibatch of $N$ image patches, we define two correlated views $\mathbf{x}_{k}^{c_1}$ and $\mathbf{x}_{k}^{c_2}$ of the same input patch $\mathbf{x}_k$ in the minibatch by randomly selecting two channels $c_1$ and $c_2$. We then promote similarity between their feature representations by minimizing the Normalized-Temperature Cross-Entropy (NT-Xent) loss \cite{sohn2016improved}, defined as:
\begin{align*}\label{eq:stageI_loss}
\ell(c_1,c_2) = \sum_{(i,j)} \sum_{k} -\log \frac{\exp(\mathrm{sim}(\mathbf{z}_{(i,j),k}^{c_1},\mathbf{z}_{(i,j),k}^{c_2})/\tau)}{\sum_{l\neq k }\exp(\mathrm{sim}(\mathbf{z}_{(i,j),k}^{c_1},\tilde{\mathbf{z}}_{(i,j),l})/\tau)}
\end{align*}
where $\mathbf{z}_{(i,j),k}^{c_1}$ is the value of $\mathbf{z}_{k}^{c_1}=g(f((\mathbf{x}_{k}^{c_1}))$ at spatial location $(i,j)$, $\tilde{\mathbf{z}}_{(i,j),l}$ is the value of $\tilde{\mathbf{z}}_{l}=g(f(\tilde{\mathbf{x}}_{l}))$ at $(i,j)$, $\tilde{\mathbf{x}}_l$ is a view of the input image $\mathbf{x}_l$ (i.e., $\tilde{\mathbf{x}}_l$ corresponds to either $\mathbf{x}_{l}^{c_1}$ or $\mathbf{x}_{l}^{c_2}$), $\mathrm{sim}(\bold{u},\bold{v})=\frac{\bold{u^{T}\bold{v}}}{||\bold{u}||||\bold{v}||}$ is the cosine similarity between the feature vectors $\bold{u}$ and $\bold{v}$, and $\tau$ is a temperature hyper-parameter which controls the rate of convergence. 
Notice that this task is applied not only to promote similarity between SAR and optical but also between different optical bands. 

Since this pretraining task is applied to the outputs of the single-channel encoder, a relatively shallow preprocessor, the feature space is still mostly affected by low-level image characteristics, as desired.

\subsubsection{CoRe -- context reconstruction to promote material features}

The second issue we address is also specific to the remote sensing scenario. 
In many remote sensing problems, such as land cover classification, the class label is mostly related to the spectral properties of the scene, and only weakly to its geometric appearance. This suggests that features representing material properties useful for land cover mapping cannot be extracted by self-supervised approaches that contrast views obtained via geometric augmentations (e.g., rotations).
For this reason, we propose \textit{CoRe} (Context Reconstruction), depicted in Fig. \ref{fig:core}: a pretext task that can be solved in a self-supervised manner and whose solution promotes features that capture material properties and thus cluster according to land cover labels.  In this pretext task, the input image is first corrupted using a given degradation process, then the network learns to reconstruct the clean image by minimizing the $\ell_1$ distance between the output of the network and the original image. In contrast to UniFeat, which only pretrains the early layers of the network, this task pretrains the entire architecture of Fig.\ref{fig:architecture}. Notice that a projection head with $C$ output channels is used during pretraining and then discarded, to be replaced with the actual head estimating the class probabilities. The input degradation process consists in the following steps: \textit{Channel dropout}, \textit{Cutout}, \textit{Gaussian blur}. 
Channel dropout randomly drops a number of input channels (putting them to 0) to promote learning features that accurately represent the spectrum, which is highly informative for material discrimination. The additional cutout and blurring degradations also add robustness, improving resilience to noise, and avoid convergence to trivial solutions, forcing the network to reason across spatial neighborhoods due to the missing regions. We remark that it might happen that different channels have different spatial resolutions (e.g., in a Sentinel 1-2 fusion problem, the multispectral bands can have resolutions of 10m, 20m or 60m, and 10m or more for SAR). In the case where all the channels at higher resolutions are dropped, the pretraining task becomes an inter-band super-resolution problem, which further promotes the emergence of features with high spatial resolution. Additionally, in a SAR-optical fusion setting, the task also requires to predict one modality from one other, further enhancing the creation of a shared feature space.

\section{Experimental Results}\label{sec:results}

\begin{figure*}
\centering
\includegraphics[width=0.16\textwidth]{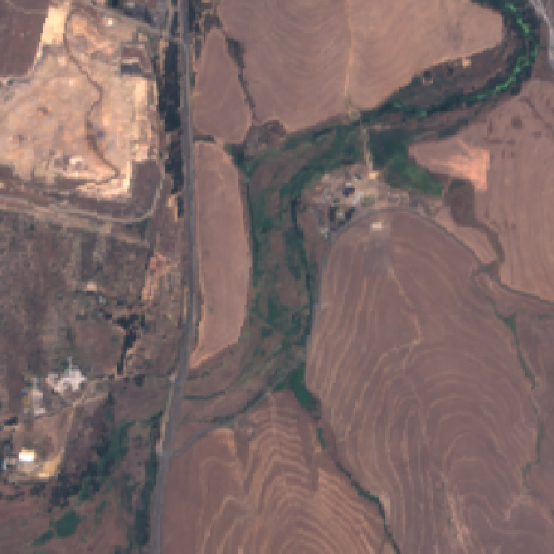}
\includegraphics[width=0.16\textwidth]{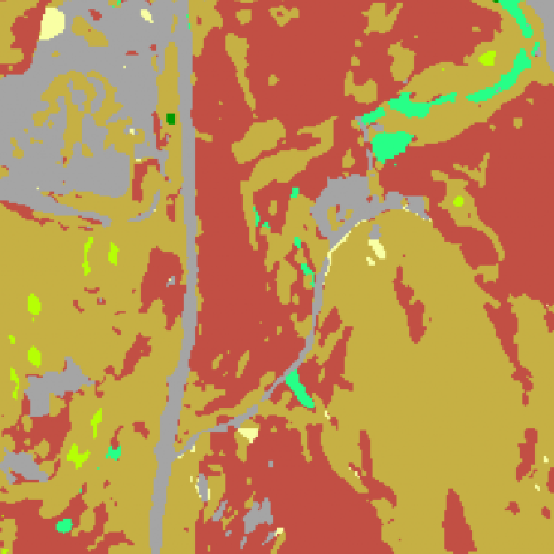}
\includegraphics[width=0.16\textwidth]{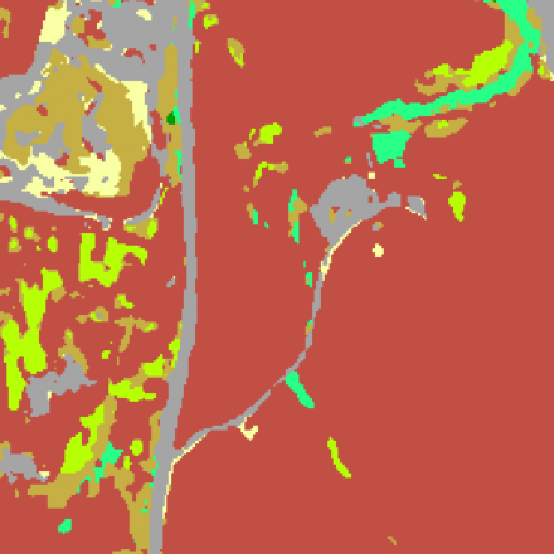}
\includegraphics[width=0.16\textwidth]{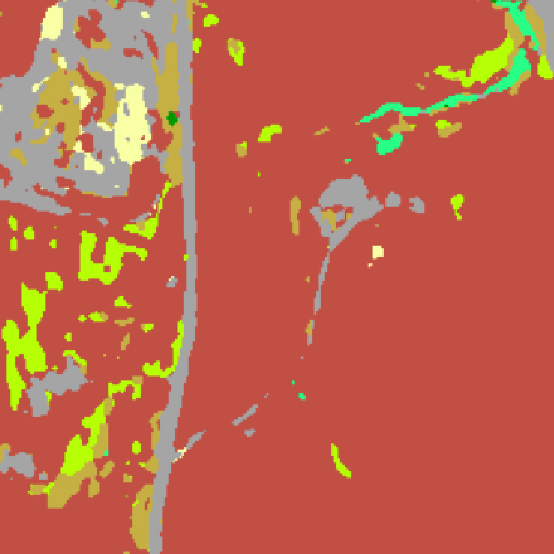}
\includegraphics[width=0.16\textwidth]{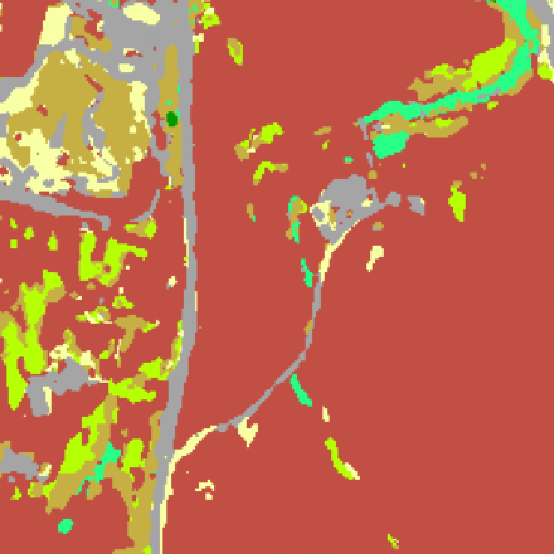}
\includegraphics[width=0.16\textwidth]{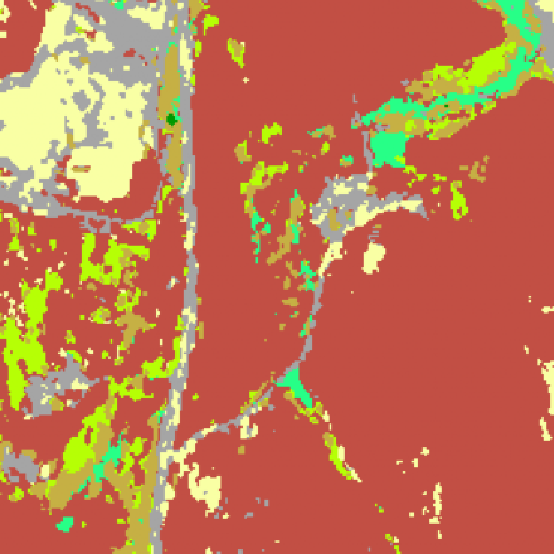}\\
\includegraphics[width=\textwidth]{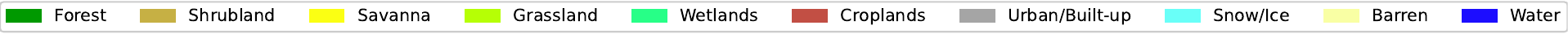}
\caption{Example of the generated result. From left to to right: Input, Baseline, ImageNet, SimSiam, SSCL (ours), ground truth.}
\label{fexample}
\vspace*{-10pt}
\end{figure*}

We test the proposed SSCL method on the dataset used for Track 2 of DFC2020 challenge \cite{9369830} organised by the Image Analysis and Data Fusion Technical Committee of the IEEE Geoscience and Remote Sensing Society, which is a subset of the SEN12MS dataset \cite{schmitt2019sen12ms}.
The input images are acquired by 2 sensors: Sentinel 1 (S1) SAR with 2 channels (VV and VH polarizations) and Sentinel 2 (S2) multispectral with 13 channels. All data are provided at a ground sampling distance equal to 10m and a fixed image size of $256\times256$ pixels.
The semantic maps have a resolution of 10m and follow a simplified version of IGBP classification scheme,  aggregated to 10 less fine-grained classes. 
We use 5128 scenes for pretraining, then the same are employed for supeervised finetuning (4128 for training, 1000 for validation). Finally, the model is tested on 986 scenes never seen before.
We use overall accuracy (OA), average accuracy (AA) and mean Intersection over Union (mIoU) as evaluation metrics.

\begin{table}[t]
\centering
 \caption{Test accuracy for the linear protocol of DeepLab at different initializations.}
 \begin{tabular}{ccccc}
   & Random init & ImageNet & SimSiam & \textbf{SSCL} \\
  \hline
  \hline
   AA & $35.1^{\pm0.1}$ & $30.9^{\pm0.3}$ &  $29.2^{\pm0.1}$  & $\bold{41.6}^{\pm0.1}$\\
  \hline
   OA & $50.1^{\pm0.1}$ & $45.4^{\pm0.3}$ &  $46.8^{\pm0.2}$  & $\bold{57.2}^{\pm0.2}$\\
  \hline
   mIoU & $19.0^{\pm0.1}$ & $15.5^{\pm0.1}$ &  $14.5^{\pm0.1}$  & $\bold{24.5}^{\pm0.3}$\\
  \hline
 \end{tabular}
 \label{taalp}
\end{table}

\subsection{Main results}\label{sec:main_results}

\begin{table}[]
\renewcommand*{\arraystretch}{1.2}
\centering
 \caption{Class-wise average and overall accuracies for a single-channel FE DeepLab with different initializations.}
 \vspace{-3pt}
  \begin{tabular}{lllll}
    & Random init. & ImageNet & SimSiam & \textbf{SSCL} \\
  \hline
  \hline
    Forest  & $64.2 ^{\pm 24}$&  $62.5 ^{\pm 17.2}$&  $\bold{76.3}^{\pm 3.1} $& $73.1^{\pm11.6}$\\
  \hline
     Shrubland & $55.4 ^{\pm 2.8}$&  $50.7 ^{\pm 3.7}$&  $52.7^{\pm 4.1}$ & $\bold{56.5}^{\pm1.7}$\\
  \hline
    Grassland  & $47.1 ^{\pm 12}$&  $46.3 ^{\pm 17.1}$&  $37.9 ^{\pm7.1}$& $\bold{54.0} ^{\pm22.0}$\\
  \hline
    Wetlands  & $\ 7.8 ^{\pm 4.8}$&  $21.6 ^{\pm 11.8}$&  $\ 5.2^{\pm1.0}$& $\bold{21.7}^{\pm 16.8}$\\
  \hline
    Croplands  & $77.5 ^{\pm 10.3}$&  $\bold{83.9}^{\pm6.4}$& $81.6 ^{\pm6.3}$& $78.2 ^{\pm 7.1}$\\
  \hline
    Urban  & $82.2 ^{\pm 2.3}$&  $77.5 ^{\pm 1.8}$&  $78.1^{\pm2.8}$& $\bold{83.1}^{\pm1.6}$\\
  \hline
    Barren  & $79.6 ^{\pm 3.1}$&  $78.3^{\pm 3.3}$&  $76.6 ^{\pm4.5}$ & $\bold{80.6}^{\pm3.7}$\\
  \hline
    Water  & $99.5^{\pm 0.1}$&  $99.3^{\pm0.1}$&  $\bold{99.6}^{\pm0.1}$& $99.3^{\pm0.3}$\\
  \hline
  \hline
    $\bold{AA}$  & $64.2^{\pm3.1}$&  $65.0^{\pm2.2}$&  $63.5^{\pm0.6}$ & $\bold{68.3}^{\pm1.2}$\\
  \hline
    $\bold{OA}$  & $67.4 ^{\pm 2.7}$ &  $69.8^{\pm1.4}$ & $67.0^{\pm0.8}$ & $\bold{71.6}^{\pm0.4}$\\
  \hline
    $\bold{mIoU}$  & $45.3 ^{\pm 3.1}$ &  $48.0^{\pm1.5}$ & $45.1^{\pm0.5}$ & $\bold{49.6}^{\pm0.8}$\\
  \hline
 \end{tabular}
 \label{taaft}
\vspace*{-15pt}
\end{table}

\begin{table}[t]
\centering
 \caption{Test accuracy of SSCL compared to the self-supervised strategy PixIF \cite{chen2021self} .}
 \begin{tabular}{ccccc}
   & \multicolumn{2}{c}{Linear Protocol} & \multicolumn{2}{c}{Finetune} \\
   & PixIF & SSCL & PixIF & SSCL \\
  \hline
  \hline
  AA & $57.0^{\pm0.4}$ & $41.6^{\pm0.1}$& $60.1^{\pm0.4}$ & $\bold{68.3}^{\pm1.2}$\\
  \hline
  OA & $63.0^{\pm0.2}$ & $57.2^{\pm0.2}$ & $65.2^{\pm0.6}$ & $\bold{71.6}^{\pm0.4}$\\
  \hline
  mIoU & $34.7^{\pm0.2}$ & $24.5^{\pm0.3}$& $38.0^{\pm0.2}$ & $\bold{49.6}^{\pm0.8}$\\
  \hline
 \end{tabular}
 \label{taacomp}
\vspace*{-20pt}
\end{table}

\begin{figure}[tb]
    \vspace*{-15pt}
    \centering
    \includegraphics[width=7cm]{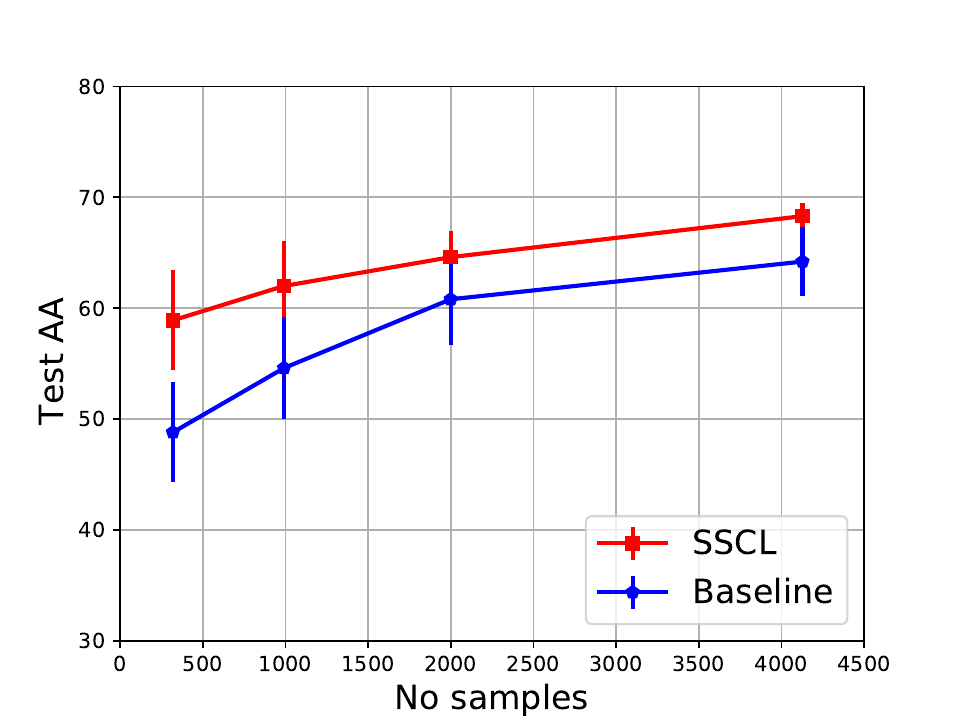}
    \caption{Test average accuracy over the training samples.}
    \label{faa}
\vspace*{-10pt}
\end{figure}

We first assess the effectiveness of the self-supervised learning stages.
The established method to evaluate this is the linear protocol, which consists in training a linear classifier on top of the network, while the weights of the neural network are frozen to the values optimized by the self-supervised pretraining. 
We compare the proposed method against a randomly initialized network with the same architecture, with respect to using classic pretraining on ImageNet and a self-supervised pretraining method which is state-of-the-art on computer vision tasks, namely SimSiam \cite{Chen_2021_CVPR}. Note that in this case we follow the standard augmentations in computer vision, i.e. geometric transformations and Gaussian blur. 
Our architecture follows the general scheme of Fig. \ref{fig:architecture}, with DeepLabv3 as state-of-the-art segmentation network. Table \ref{taalp} reports the results in terms of AA , OA and mIoU.
We can observe that the pretraining on ImageNet and SimSiam are not effective, confirming the domain gap between traditional whole--image classification in computer vision and land cover classification.
On the other hand, the proposed method shows higher accuracy than random initialization, confirming our conjecture that the proposed self-supervised tasks are able to better capture the information related to material properties.

We then focus our attention on evaluating the finetuning performance (Table \ref{taaft}), i.e., when the entire pretrained model is optimized using the available labels. We compare against the same initialization schemes of the previous experiment.
It can be noticed that the proposed approach is the only one that is able to significantly improve over random initialization.

These results suggest that the proposed method is highly effective at improving the performance of end-to-end deep learning models for land cover classification when SAR and multispectral data are jointly used. 
A qualitative comparison is shown in Fig. \ref{fexample}, which shows some examples of predicted maps obtained using the different methods considered in the evaluation. We can observe that the proposed SSCL is able to segment finer details than existing methods. Also notice that, according to visual inspection, in some cases, SSCL seems to be even more accurate than the ground truth due to mislabeling issues in the dataset, especially for similar classes such as Shrubland, Grassland and Forest.

Finally, Table \ref{taacomp} reports a comparison with the recently proposed self-supervised contrastive learning method PixIF \cite{chen2021self}. We retrained PixIF to match our experimental setting using the authors' code. We can notice that PixIF is very effective in the self-supervised setting, outperforming SSCL on the linear protocol. However, SSCL is superior in the semi-supervised setting when finetuned using labels (even a small amount, as in Fig. \ref{faa}). We thus consider PixIF complementary to our work. 

\vspace{-3pt}
\subsection{Analysis and ablation experiments}

First, we are interested in evaluating the performance improvements provided by SSCL under label scarcity. Fig. \ref{faa} shows the test AA reached when finetuning with a limited number of labels. It is interesting to notice that SSCL with just 1000 samples provides comparable performance to a randomly initialized network trained on 4128 samples.

\begin{figure}[tb]
\centering

\begin{subfigure}[b]{\columnwidth}
\centering
\includegraphics[width=0.31\textwidth]{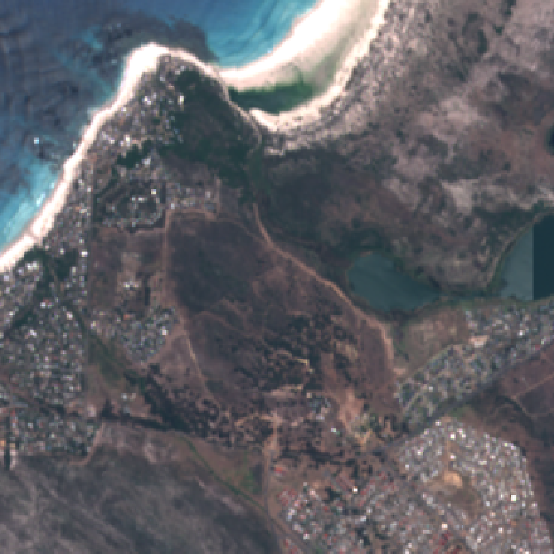}
\includegraphics[width=0.31\textwidth]{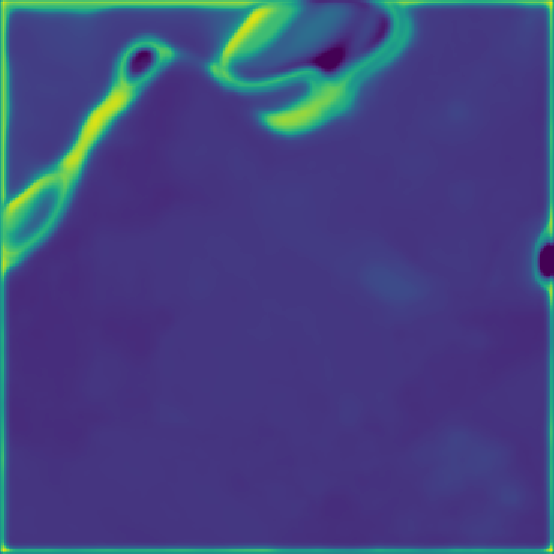}
\includegraphics[width=0.31\textwidth]{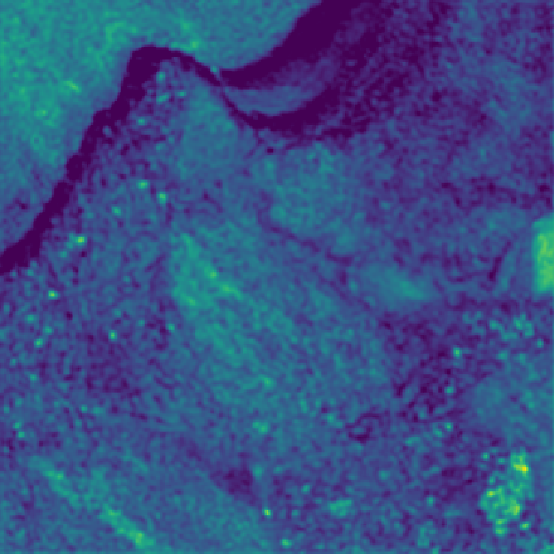}
\end{subfigure}
\caption{Spatial resolution of a feature map for SimSiam (centre) and the proposed SSCL (right). Notice the significantly higher spatial resolution of SSCL.}
\label{ffeatu}
\vspace*{-12pt}
\end{figure}

Then, we want to validate our claim that SSCL is able to capture high-resolution features with its self-supervised tasks. Fig. \ref{ffeatu} shows some representative feature maps from the last network layer and compares them between the SSCL and SimSiam. We can immediately notice that the spatial resolution of the feature maps obtained with the proposed method is much higher than SimSiam, and finer details are preserved. This correlates with the finer segmentation maps in Fig. \ref{fexample} and could be explained by the fact that the pretraining reconstruction task promotes high-resolution solutions since it has to solve problems that amount to super-resolution/deblurring (e.g., when the highest-resolution channels are dropped) or inpainting, thus heavily relying on fine spatial clues.

In the following we report ablation experiments to validate the contributions of various proposed components.
Firstly, the importance of the general archicture based on single-channel feature extractors is assessed in Table \ref{taa_monob}. The results show that Deeplab with a single-channel feature extractor outperforms a standard Deeplab with the first layer merging all the input channels and comparable number of parameters. Note that, for a fair comparison, we show the models without any pretraining in the first two columns and the model with SSCL pretraining in the last column. 
In addition, the same table shows the performance difference when those models do or do not process SAR images or have only SAR images, in order to evaluate how well they are able to exploit this information and fuse it with multispectral images.
We can observe that effective fusion between SAR and multispectral information is achieved by the proposed method.

Finally, in Table \ref{taaprep} we test the effect of UniFeat. In particular, we are interested in showing that it can perform more than a simple denoising of the SAR input and without manual design of the preprocessing function. We substitute UniFeat with a conventional despeckling algorithm (SAR BM3D) and notice that we obtain similar results. However, when we use the SSCL including UniFeat and the manual preprocessing, we observe an improvement, confirming that UniFeat acts not only as a denoiser of SAR images but as a more complex regularizer reducing intra-class variance across modalities.

\begin{table}[tb]
\centering
 \caption{Test average accuracy for DeepLab with or without single-channel FE and with or without SAR images.}
 \begin{tabular}{ lccc  }
      & Std Deeplab & Single-ch. FE  & SSCL    \\
  \hline
  \hline
  with SAR    & $61.7^{\pm2.0}$ &  $64.2^{\pm3.1}$  & $\boldsymbol{68.3}^{\pm1.2}$ \\
  \hline
  w/o SAR    & $59.7^{\pm1.9}$ &  $59.5^{\pm2.3}$   & $\boldsymbol{67.6}^{\pm1.6}$ \\
  \hline
  only SAR    & $54.8^{\pm1.2}$ &  $55.9^{\pm0.6}$   & $\boldsymbol{56.3}^{\pm0.3}$ \\
  \hline
 \end{tabular}
 \label{taa_monob}
\vspace*{-4pt}
\end{table}

\begin{table}[t]
\setlength{\tabcolsep}{8pt}
\centering
 \caption{Test average and overall accuracy of our SSCL with and without UniFeat and a manual preprocessing}
 \begin{tabular}{lcccc}
    & CoRe  & Preproc + CoRe & SSCL & Preproc + SSCL\\
  \hline
  \hline
  AA    &  $67.4^{\pm1.3}$  & $ 68.3^{\pm1.5}$ & $ 68.3^{\pm1.2}$ & $ \boldsymbol{69.4}^{\pm0.7}$\\
  \hline
  OA    &  $70.2 ^{\pm 0.9}$ & $71.0^{\pm 0.9}$ & $ 71.6^{\pm0.4}$& $ \boldsymbol{72.3} ^{\pm 0.6}$\\
  \hline
 \end{tabular}
 \label{taaprep}
\vspace*{-10pt}
\end{table}

\vspace{-3pt}
\section{Conclusions}
\vspace{-3pt}
In this paper, we proposed a framework for self-supervised pretraining of deep neural networks for the task of semi-supervised land cover classification. We showed how the proposed method is effective at jointly processing images from multiple sensing modalities, such as SAR and multispectral.
\vspace{-3pt}


\end{document}